\documentclass[11pt]{article}

\usepackage{graphicx}
\begin{document}
\begin{center}
%\title
{\Large\bf On the long-range correlations in hadron-nucleus 
collisions}\\

\vskip 0.5cm

%\author
{N.Armesto$^{a}$, M.A.Braun$^{b}$ and C.Pajares$^{a}$}
%\thanks{^{a)} Dep. Fisica de Particulas, Universidad de Santiago, Spain,
\vskip 0.3cm

$^a$
Departamento de F\'{\i}sica de Part\'{\i}culas and IGFAE,\\
Universidade de Santiago de
Compostela,
 15782 Santiago de Compostela, Spain
\vskip 0.2cm

$^{b}$ Department of  High-Energy Physics,
St. Petersburg University,\\ 198504 St. Petersburg, Russia
\end{center}
\date{}
\pagestyle{myheadings}
\def\beq{\begin{equation}}
\def\eeq{\end{equation}}
\def\noi{\noindent}
\def\bk{{\bf k}}
\def\tka{\tilde{k}_1}
\def\tkb{\tilde{k}_2}
\def\tr{\tilde{r}}
\def\phid{\phi^{\dagger}}
\def\psid{\psi^{\dagger}}
\def\by{\bar{y}}
\def\ba{\bar{\alpha}}

%\begin{document}
%\maketitle
\medskip
\vspace{1 cm}

{\bf Abstract}

Long-range correlations between multiplicities in different rapidity
windows
in hadron-nucleus collisions are analyzed.
After recalling the standard results in the probabilistic model, we
study them in the framework of perturbative QCD. Considering interacting BFKL
pomerons in the form of fan diagrams coupled to a dilute projectile, analytic
estimates are done for very large rapidities. The correlation strength results
weakly depending on energy and centrality or nuclear size, and generically
greater than unity. Finally, we turn to the Color Glass Condensate framework.
For a saturated projectile and considering the most feasible
experimental situation of forward and backward rapidity windows symmetric
around the center-of-mass, the resulting correlation
strength turns out to be larger than unity and
shows a non-monotonic behavior with increasing energy, first increasing and
then decreasing to a limiting value. Its behavior with
increasing centrality or nuclear size depends on the considered
rapidity
windows.

\vspace{1 cm}

\section{Introduction}
Long-range correlations have been attracting much attention since long ago
in the region of both low and high transverse momenta of secondaries.
At low momenta the color string picture \cite{capella}
with fusion and percolation effects \cite{perc}
has been extensively applied   \cite{amelin,kolevatov}. In the 
semihard region
the Color Glass Condensate (CGC) picture has lately been used
\cite{levin,AMcLP}.
One can separate  long-range correlations into a contribution from purely
hadronic
collisions and another coming from the effects generated by the heavy nucleus
target or/and projectile.  Obviously the first contribution can hardly be
treated in
a more or less rigorous theoretical framework due to the essentially
non-perturbative structure
of the hadron.  Heavy nucleus collisions, on the other hand, present more
opportunities in
this sense, due to their comparatively simple structure in terms of
constituent nucleons.
Single inclusive cross-sections with participation of nuclei can be easily
found even in the
framework of interacting pomerons, both of the old local type and
of the sophisticated Balitsky-Fadin-Kuraev-Lipatov (BFKL)
type. In the limit of very heavy nuclei they are expressed in terms of the
sum of the
corresponding pomeron fan diagrams. However long-range correlations require
also knowledge
of the double inclusive cross-section, for which the situation is more
complicated. In the
purely eikonal approach they can also be easily calculated from the known
single and double
elementary (hadron-nucleon) inclusive cross-sections. However with
interacting pomerons this is only possible for the hadron-nucleus case. Double
inclusive cross-section for nucleus-nucleus scattering mediated by
interacting pomerons involves a complicated set of diagrams,
exact summation of which does not look feasible. So hadron-nucleus
collisions present
a subject better suited for  theoretical discussion of long-range
correlations
in the nuclear background.

In this paper we study long-range correlations in hadron-nucleus
collisions in the hard domain.
We shall use two different approaches to this problem, which treat
different range of
energies.  At very high (asymptotic) energies we shall rely on the
perturbative Quantum Chromodynamics (QCD),
which predicts that the interaction is mediated by  interacting BFKL
pomerons while treating the hadron as a dilute object through the whole
evolution. At smaller
energies, when evolution of the gluon density is not complete, we shall
apply the Color Glass Condensate approach, in which the fast nucleus is
represented by a strong classical gluon field
\cite{mclerran}. This approach has  been lately used for a qualitative study of
correlations
in nucleus-nucleus collisions with promising results  \cite{AMcLP}.

The paper is organized as follows. In the next section, to have a basis
for comparison and discussion, we shall recall some basic predictions for
the long-range correlations in $hA$ collisions, which follow from the
straightforward probabilistic approach. Section 3 is dedicated to the
perturbative QCD approach at very high energies. Section 4 presents some
qualitative predictions
from the Color Glass Condensate. Finally we discuss our results in
the Conclusions.

\section{Probabilistic treatment}
In this section we shall study long-range correlations in $hA$ collisions, as
they follow from purely
probabilistic considerations. This approach is realized in the
Glauber-eikonal description of
$hA$ interactions and also reproduced in Regge approach with non-interacting
pomerons.
Our basic instrument will be the standard Glauber expression for the
cross-section
$\sigma_n$ for $n$ inelastic interactions of the projectile inside the nucleus:
 \beq
\sigma_n=C_A^n\Big(\sigma T(b)\Big)^n\Big(1-\sigma T(b)\Big) ^{A-n}.
\eeq
Here $\sigma$ is the elementary (hadron-nucleon) cross-section, and
$T(b)$ is the standard nuclear profile function normalized to unity.
From this expression we immediately derive expressions for the single and
double
inclusive cross-sections for $hA$ collisions, $J_1$ and $J_2$ respectively.
At fixed impact parameter $b$
\beq
J_1(y,k)=Aj_1(y,k)T(b)
\eeq
and
\beq
J_2(y_1,k_1;y_2,k_2)=Aj_2(y_1,k_1;y_2,k_2)T(b)+A(A-1)j_1(y_1,k_1)j_1(y_2,k_2)T^2
(b).
\eeq
Here and in the following we denote by small letters the quantities which
refer to the elementary $hN$ collision. So $j_1$ and $j_2$ are the single and
double inclusive cross-sections for 
hadron-nucleon collisions; $y$ and $k$ denote the rapidity and transverse
momentum of the
produced particle.

To pass to the corresponding multiplicities we have to integrate over $b$
and divide by the
total inelastic $hA$ cross-section $\Sigma^{in}$. To do this we have to
choose a form of the profile function
$T(b)$. We use the simplest choice of the constant nuclear density within a
sphere of
radius $R_A=A^{1/3}R_0$, which gives
\beq
T(b)=\frac{2\sqrt{R_A^2-b^2}}{V_A},
\label{profile}
\eeq
where $V_A$ is the nuclear volume. With this  profile function we find
\beq
\int d^2b T^2(b)=\frac{9}{8}\frac{1}{\pi R_A^2}.
\eeq
For large $A$, with a good precision, the inelastic $hA$ cross-section
$\Sigma^{in}=\pi R_A^2$,
so that we finally find the single and double $hA$ multiplicities, $M_1$ and
$M_2$, respectively as
\beq
M_1(y,k)=A^{1/3}m_1(y,k)
\label{nucm1}
\eeq
and
\beq
M_2(y_1,k_1;y_2,k_2)=A^{1/3}m_2(y_1,k_1;y_2,k_2)+
\frac{9}{8}A^{2/3}\mu_1(y_1,k_1)m_1(y_2,k_2).
\label{nucm2}
\eeq

With these expressions we can pass to correlations. The strength of
long-range correlations
in nuclear collisions is standardly determined by the coefficient
\beq
B=\frac{ \langle N_FN_B\rangle -\langle N_F\rangle \langle N_B\rangle
}{\langle N_F^2\rangle -\langle N_F\rangle ^2},
\label{nuclb}
\eeq
where $N_F$ and $N_B$ are the numbers of particles produced in two
rapidity windows. separated by a sufficiently large rapidity interval
('forward' and 'backward'). Note that in the asymmetric hadron-nucleus case,
there is another correlation strength defined with $\langle N_F^2\rangle
-\langle N_F\rangle ^2$ in
the denominator. As to the particle transverse momenta, they
may be both taken integrated over the whole phase space
or restricted to specific parts of it (even practically fixed).  This
circumstance plays no
role for the following derivation. The average numbers which figure in
(\ref{nuclb}) are expressed
via the multiplicities as follows
\beq
\langle N_{F(B)}\rangle =\int d\tau^{F(B)}M_1(y,k),
\label{nucsing}
\eeq
\beq
\langle N_{F}N_{F(B)}\rangle =\int d\tau^F_1d\tau^{F(B)}_2M_2(y_1,k_1;y_2,k_2),
\label{nucdoub}
\eeq
where $d\tau^F$ and $d\tau^b$ denote integration over $y$ and $k$ in the
forward and backward windows.

Similar quantities for the elementary $hN$ collisions will be denoted by
small letters. So for
$hN$ collisions the correlation coefficient is determined as
\beq
b=\frac{ \langle n_Fn_B\rangle -\langle n_F\rangle \langle n_B\rangle
}{\langle n_F^2\rangle -\langle n_F\rangle ^2}\,,
\label{elemb}
\eeq
where the averages are defined as in (\ref{nucsing}) and (\ref{nucdoub})
with multiplicities
$\mu_1$ and $\mu_2$. Using relations (\ref{nucm1}) and   (\ref{nucm2}) we can
express averages for the nucleus target via the same quantities on the
nucleon target
to obtain:
\beq
B=\frac{ \langle n_Fn_B\rangle +\frac{1}{8}A^{1/3}\langle n_F\rangle \langle
n_B\rangle }
{\langle n_F^2\rangle +\frac{1}{8}A^{1/3}\langle n_F\rangle ^2}.
\eeq
If we define the dispersion squared in the forward window for the
elementary collisions  as
\beq
d^2=\langle n_F^2\rangle -\langle n_F\rangle ^2,
\eeq
then we find
\beq
B=\frac{bd^2+\Big(\frac{1}{8}A^{1/3}+1\Big)\langle n_F\rangle \langle
n_B\rangle }
{d^2+\Big(\frac{1}{8}A^{1/3}+1\Big)\langle n_F\rangle ^2}.
\label{nucbfin}
\eeq
For symmetric windows (relative to $hN$ collisions) we have $\langle n_F\rangle
=\langle n_B\rangle $ so that
(\ref{nucbfin}) simplifies to
\beq
B=\frac{ bd^2+\Big(\frac{1}{8}A^{1/3}+1\Big)\langle n_F\rangle ^2}
{d^2+\Big(\frac{1}{8}A^{1/3}+1\Big)\langle n_F\rangle ^2}.
\label{nucbfin1}
\eeq
If the dispersion squared  is much smaller than the particle number
squared, which is expected
for large enough energies and windows, we find an approximate expression
\beq
B\simeq 1-\frac{d^2}{\langle n_F\rangle
^2}(1-b)\frac{1}{1+\frac{1}{8}A^{1/3}}\,.
\eeq

As a result we find that in the theoretical limit $A\to\infty$ the nuclear
coefficient $B$
tends to unity, the value of the elementary coefficient $b$ having no
importance.
So, in this limit, long-range correlations are
exclusively a consequence of the nuclear environment. Of course for realistic
nuclei the term
$(1/8)A^{1/3}$ is not large but
smaller than unity so that  the coefficient $B$ results noticeably smaller
than unity.
However we also see that its value only weakly depends on $A$
and is mainly determined by the relative dispersion $d/\langle n_F\rangle $ in elementary
collisions.
 All these effects are due to the presence of the second term in the
nuclear multiplicity $M_2$, Eq. (\ref{nucm2}). Note that the limiting case
$A=1$ is achieved through the substitution
$A^{1/3}/8\to -1$ in (\ref{nucbfin}),
so that
the second terms in both the numerator and denominator of this equation
vanish
and $B$ passes into $b$.

\section{Long-range correlations in high-energy $hA$ collisions in the
perturbative QCD}
\subsection{Formalism}
In the perturbative QCD at high energies the interaction between the
incoming hadron and the nuclear target is realized by an exchange of BFKL
pomerons, which interact between themselves via the triple pomeron
vertex. For a heavy nucleus target with
$A\gg 1$ and a point-like projectile this interaction is described by a set of
pomeron fan diagrams, which are summed by the
non-linear Balitsky-Kovchegov equation \cite{bal,kov, braun1}.  If we denote
this sum for a fixed
impact parameter $b$, zero total gluon momentum and intergluon transverse
distance $r$ as $\Phi(y,r,b)$ then  function $\phi(y,r,b)=\Phi(y,r,b)/(2\pi
r^2)$, transformed into momentum space, satisfies the equation
\beq
\frac{\partial\phi(y,q,b)}{\partial \bar{y}}=-H\phi(y,q,b)-\phi^2(y,q,b),
 \eeq
where $H$ is the BFKL Hamiltonian (see e.g \cite{lipatov}) and
 $\bar{y}=\bar{\alpha}y$ is a 
rescaled rapidity with the standard
notation
$\bar{\alpha}=\alpha_sN_c/\pi$.
The physical meaning of $\phi$ is provided by its relation to the gluon
density in the nucleus:
\beq
\frac{d\Big(xG(x,k^2,b)\Big)}{d^2kd^2b}=
\frac{N_c^2}{2\pi^3}\frac{1}{\bar{\alpha}}h(y,k,b),
\eeq
where
\beq
h(y,k,b)=k^2\nabla_k^2\phi(y,k,b)
\eeq
and $y=\ln (x_0/x)$ with $x_0$ some initial value of $x$, usually taken $\sim
0.01$.
Function $h$ satisfies a normalization condition
\beq
\int \frac{d^2k}{2\pi k^2}h(y,k,b)=1.
\label{norm}
\eeq
Numerical calculations \cite{armesto} show that starting from
$\bar{y}\simeq 2$
it develops a scaling structure
\beq
h(y,k,b)=h(\xi)=0.295\,\exp (-\xi^2/3.476), \ \ \xi=\ln\Big(k/Q_s(y,b)\Big),
\label{scale}
\eeq
where $Q_s(y,b)$ is the so-called saturation momentum, which grows
exponentially with $y$:
\beq
Q_s(y,b)\simeq a\Big(AT(b)R_0^2\Big)^{2/3}\frac{e^{\Delta_1y}}{\sqrt{y}}\,.
\eeq
 $\Delta_1=(2.0\div 2.3)\ba$ and $a$ is a numerical constant\footnote{This
dependence
on impact parameter and nuclear size was obtained \cite{armesto}
for a realistic profile
function. Also the value of $\Delta_1$ depends on the rapidity window of the
fit, and is slightly smaller than the asymptotic theoretical expectation,
$\Delta_1=2.44 \ba$. See detailed discussions of these aspects in
e.g. \cite{Albacete:2004gw} where a dependence
$\propto A^{1/3}$ is obtained for a cylindrical nucleus. None of these
considerations alter the conclusions obtained in this Section.}.

An immediate physical application of this framework is to deep inelastic
scattering (DIS), with a
highly virtual photon as a projectile.  A hadronic analogue of this may be
the 'onium', that is a quark-antiquark system of a very short dimension.
Realistic hadronic projectiles are not point-like and do not allow for the
perturbative treatment. So application of the fan diagram approach to their
interaction
is strictly speaking not very well justified.  For this reason
 in the following we have
to keep in mind
an  approximate character of our derivation, which assumes that, as in DIS,
the projectile
hadron interacts with the pomeron only once.

Under this approximation the total hadron-nucleus cross-section is given by
\beq
\Sigma(Y)=2\int d^2bd^2\rho(r)\Phi(Y,r,b),
\eeq
where $\rho(r)$ is the color density of the projectile and $Y$ is the
overall rapidity. It may be illustrated graphically as
shown in Fig. 1, where the circle with the attached line correspond to
function $\Phi$.
For a normalizable density $\rho(r)$ and finite nucleus (with the profile
function
(\ref{profile})) cross-section $\Sigma=\Sigma^{in}$ turns out to be purely
geometric: $\Sigma=2\pi R_A^2$.
%%%%%%%%%%%%%%%%%%%%%%%%%%%%%%%%%%%%%%%%%%%%%%%%%%%%%
\begin{figure}[ht]
\centering
\includegraphics[width=1.5cm]{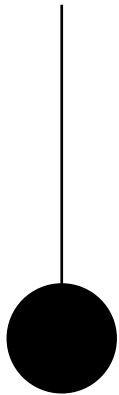}
\caption{Diagram contributing to the total $hA$ cross-section.}
\label{Fig1}
\end{figure}
%%%%%%%%%%%%%%%%%%%%%%%%%%%%%%%%%%%%%%%%%%%%%%%%

\subsection{Single inclusive cross-sections}
 The single inclusive cross-section is given by the two diagrams shown in
Fig. 2 $a$ and $b$. They correspond to production of the observed particle
either from the pomeron which couples to the
projectile or from the  vertex of its splitting into a pair of pomerons
\cite{braun2, tuchin}.
All other possibilities
are canceled by the Abramovsky-Gribov-Kancheli cutting rules  \cite{AGK}.
%%%%%%%%%%%%%%%%%%%%%%%%%%%%%%%%%%%%%%%%
\begin{figure}[ht]
\centering
\includegraphics[width=8cm]{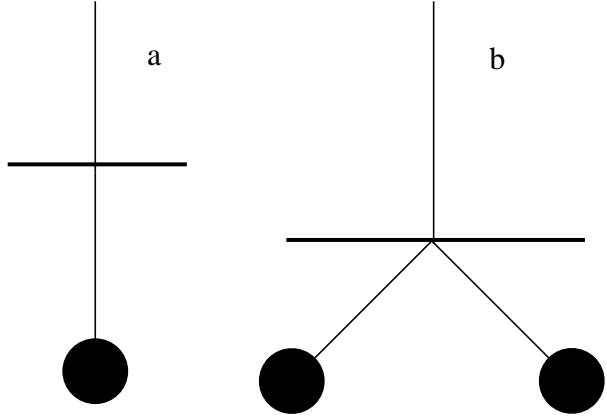}
\caption{Diagrams contributing to the single inclusive  $hA$ cross-section.}
\label{Fig2}
\end{figure}
%%%%%%%%%%%%%%%%%%%%%%%%%%%%%%%%%%%%%%%%%%%%%%%%%%%
The sum of these two contributions leads to the single inclusive
cross-section at fixed $b$
\cite{braun3}
\beq
J_1(y,k,b)=\frac{4\pi\ba}{k^2}\int 
d^2re^{ikr}\nabla^2P(Y-y,r)\nabla^2\Big(2\Phi(y,r,b)-
\Phi^2(y,r,b)\Big),
\eeq
where $P(Y-y,r)$ is the pomeron coupled to the projectile:
\beq
P(Y-y,r)=\int d^2r'\rho(r')G(Y-y,r',r),
\eeq
with the nucleus at $y=0$ and the dilute projectile at $y=Y$,
and $G$ is the BFKL forward Green function  \cite{lipatov}.
Performing the differentiations and passing to momentum space one
obtains
\beq
J_1(y,k,b)=\frac{4\pi\ba}{k^2}\int d^2q h^{(0)}(Y-y,q)w(y,k-q,b).
\eeq
Here $h^{(0)}(y,q)$ is a function analogous to $h(y,q,r)$ for the pomeron,
that is a Fourier transform of $\nabla^2P(y,r)/(2\pi)$. Function $w(y,q,b)$
is defined via $h(y,q,b)$:
\beq
w(y,q,b)=\frac{q^2}{2\pi}\int d^2q_1
\frac{h(y,q_1,b)h(y,q-q_1,b)}{q_1^2(q-q_1)^2}.
\eeq
It satisfies the same normalization condition (\ref{norm}) and has the same
scaling property
(\ref{scale}), although with a shifted maximum and slope in its  $\xi$
dependence:
\beq
w(y,k,b)=0.358\,\exp \Big(-0.402(\xi-0.756)^2\Big).
\label{scalew}
\eeq
 Of course the shift in  the maximum
corresponds to a universal enhancement of the value of the saturation momentum.

Numerical calculations of the single inclusive cross-section along these
formulas were
performed in  \cite{braun3}. However such calculations for the double
inclusive cross-sections
look very difficult, so that we shall instead use analytical estimates
obtained in the asymptotic
regime when both $Y-y$ and $y$ are large. Then we can use the well known
asymptotic expressions for the BFKL Green function  to obtain an explicit
expression for function $h^{(0)}$.
Repeating the calculations done in \cite{braun2} we obtain
\beq
J_1(y,k,b)=\frac{8\ba}{k^2}R_Pe^{\Delta (Y-y)}\sqrt{\frac{\pi}{\beta(Y-y)}}
F(y,k,b),
\eeq
where $\Delta= 4\ln 2\ba$ is the BFKL intercept,  $\beta=14\zeta(3)\ba$,
$R_P$ is
the radius of the projectile and
\beq
F(y,k,b)=\int\frac{d^2q}{2\pi q}w(y,k-q,b)=Q_s(y,b)f(\hat{k}), \ \
\hat{k}=\frac{k}{Q_s(y,b)}.
\label{ffunction}
\eeq
Function $f(\hat{k})$ obviously reduces to a constant $f(0)$ when $k\ll Q_s$. In
the opposite case
when $k\gg Q_s$ it behaves as $\lambda/\hat{k}$.  Numerical calculations give
values for $f(k)$ illustrated in Fig. 3 with
\beq
f(0)=3.97, \ \  \lambda=54.6\,.
\eeq
%%%%%%%%%%%%%%%%%%%%%%%%%%%%%%%%%%%%%%%%%%%%%%%%
\begin{figure}[ht]
\centering
\includegraphics[width=8cm]{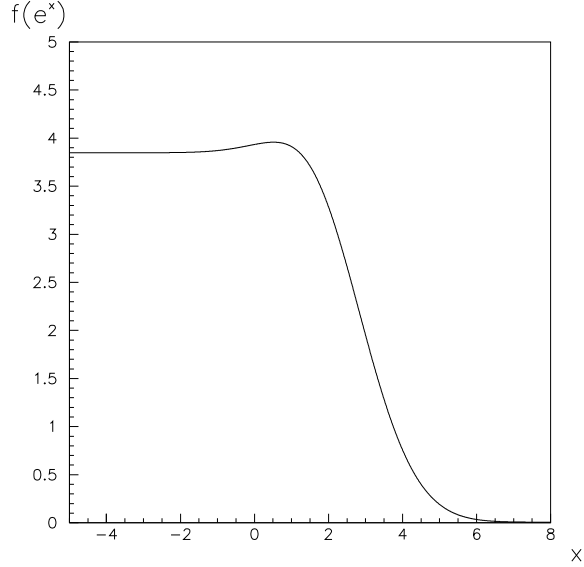}
\caption{Function $f(e^x)$.}
\label{Fig3}
\end{figure}
%%%%%%%%%%%%%%%%%%%%%%%%%%%%%%%%%%%%%%%%%%%%%%%%%%%%%%%%
So for momenta considerably below the saturation momentum we find
\beq
J_1(y,k,b)=\frac{8\ba}{k^2}R_Pf(0)e^{\Delta
(Y-y)}\sqrt{\frac{\pi}{\beta(Y-y)}}
Q_s(y,b)
\eeq
and for momenta considerably above the saturation momentum
\beq
J_1(y,k,b)=\frac{8\ba}{k^3}R_P\lambda e^{\Delta
(Y-y)}\sqrt{\frac{\pi}{\beta(Y-y)}}
Q^2(y,b)\,.
\eeq

The found cross-section grows exponentially with the overall rapidity $Y$,
which just reflects the growth of the pomeron directly coupled to the
projectile.
One expects that for an extended projectile this growth would be finally
tempered  when more than one pomeron are coupled to it (see Fig. 4 ),
contributions which are
damped by powers of the small coupling constant for a point-like projectile.
As we shall see,  in the expression for the correlation coefficient the
growing factors cancel,
so that the resulting coefficient depends on $Y$ only weakly. For this
reason we may
hope that our formulas remain applicable also for realistic hadrons.
%%%%%%%%%%%%%%%%%%%%%%%%%%%%%%%%%%%%%%%%%%%%%%%%%%%%
\begin{figure}[ht]
\centering
\includegraphics[width=2cm]{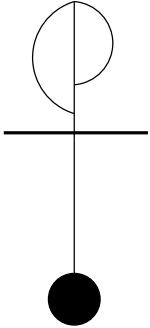}
\caption{Diagrams not taken into account for the total cross-section for a
point-like projectile.}
\label{Fig4}
\end{figure}
%%%%%%%%%%%%%%%%%%%%%%%%%%%%%%%%%%%%%%%%%%%%%%%%%%%%
Another interesting feature is the peculiar dependence on the nuclear
factor $\Big(AT(b)\Big)^p$, with $p=2/3$
 for $k\ll Q_s$ and $p=4/3$ for $k\gg Q_s$,
which is of primary importance for the correlations.
Integration over $b$ and division by $\Sigma$ leads to multiplicities in
these two regions of $k$.
For $k\ll Q_s$
\beq
M_1(y,k)=A^{2/9}\frac{3}{8}\gamma
\frac{8\ba}{k^2}R_Pf(0)e^{\Delta Y-\epsilon y}\sqrt{\frac{\pi}{\beta
y(Y-y)}},
\label{mlow}
\eeq
where
\beq
\gamma=a\Big(\frac{9}{4\pi^2}\Big)^{1/3}
\eeq
and $\epsilon=\Delta-\Delta_1$,
and for $k\gg Q_s$
\beq
M_1(y,k)=A^{4/9}\frac{3}{10}\gamma^2
\frac{8\ba}{k^3}R_P\lambda e^{\Delta Y-\epsilon_1y}
\sqrt{\frac{\pi}{\beta y^2(Y-y)}},
\label{mhigh}
\eeq
with $\epsilon _1=\Delta-2\Delta_1$.

\subsection{Double inclusive cross-section and correlations}
The double inclusive cross-section is described by a set of diagrams 
 shown in
Fig. 5 (a few evident additional diagrams are not shown).
 They are quite complicated, especially 
since the cut vertex
appearing in the diagram Fig. 5 $f$ is different from the uncut one
\cite{braun4}. As mentioned in the Introduction, a detailed calculation of
all the contributions does not seem very realistic. However at high values
of all rapidity distances,
$Y-y_1, Y-y_2, y_1,y_2\gg 1$, of all the contributions the dominant ones
correspond to Figs. 5 $c$-$e$, in which the upper vertex can have
rapidities up to $Y$, so that the two pomeron
lines below give the maximally growing exponential factor
$\exp\Big(\Delta(2Y-y_1-y_2)\Big)$. The relative weights of all other contributions
is exponentially damped.
%%%%%%%%%%%%%%%%%%%%%%%%%%%%%%%%%%%%%%%%%%%%%%%%%%%%%%%%%%%%%%%%%%%%
\begin{figure}[ht]
\centering
\includegraphics[width=12cm]{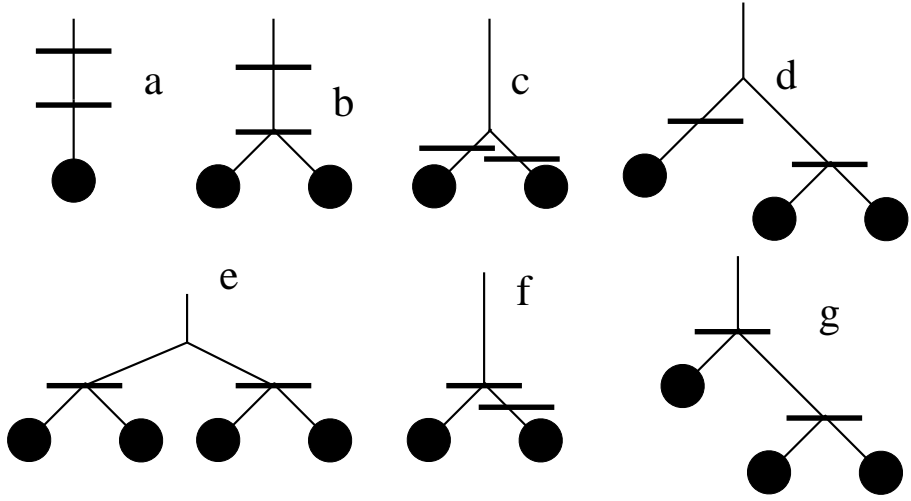}
\caption{Diagrams contributing to the double inclusive cross-section in
perturbative QCD.}
\label{Fig5}
\end{figure}
%%%%%%%%%%%%%%%%%%%%%%%%%%%%%%%%%%%%%%%%%%%%%%%%%%%%%%%%%%%%%%%%%%%%
The study of contributions from all the diagrams of Fig. 5 $c$-$e$ can be
done by the same method which was used in \cite{braun2} for the diagram of
Fig. 5 $c$.  Contribution of
the diagrams 5. $d$ and $e$   can be taken into account by changing function
$h(y,q,r)$ to $w(y,q,h)$
in all the formulas. In this way we obtain the double inclusive
cross-section as
\[
J_2(y_1,k_1;y_2,k_2,b)=
\]\beq
\frac{4}{\ln 2}\frac{\ba^2}{k_1^2k_2^2}\langle r^2\rangle _P
e^{\Delta(2Y-y_1-y_2)}\sqrt{\frac{\pi}{\beta(2Y-y_1-y_2)}}F(y_1,k_1,b)F(y_2,k_2,
b),
\eeq
where $F$ is the same function (\ref{ffunction}).
Further simplifications, similarly to the case of single inclusive
cross-sections, can be made
when $k_1$ and $k_2$ are either much smaller than the saturation momentum
or much larger than it.
In the case when both $k_1\ll Q_s$ and $k_2\ll Q_s$
\beq
J_2=\frac{4}{\ln 2}\frac{\ba^2}{k_1^2k_2^2}\langle r^2\rangle _Pf^2(0)
e^{\Delta(2Y-y_1-y_2)}\sqrt{\frac{\pi}{\beta(2Y-y_1-y_2)}}Q_s(y_1,b)Q_s(y_2,b).
\eeq
If both $k_1\gg Q_s$ and $k_2\gg Q_s$ then
\beq
J_2=\frac{4}{\ln 2}\frac{\ba^2}{k_1^3k_2^3}\langle r^2\rangle _P\lambda^2
e^{\Delta(2Y-y_1-y_2)}\sqrt{\frac{\pi}{\beta(2Y-y_1-y_2)}}Q_s^2(y_1,b)Q_s^2(y_2,
b).
\eeq
Finally in the case $k_1\gg Q_s$ and $k_2\ll Q_s$ we find
\beq
J_2=\frac{4}{\ln
2}\frac{\ba^2}{k_1^3k_2^2}\langle r^2\rangle _P\lambda f(0)
e^{\Delta(2Y-y_1-y_2)}\sqrt{\frac{\pi}{\beta(2Y-y_1-y_2)}}Q_s^2(y_1,b)Q_s(y_2,b).
\eeq
Integration over $b$ and division by $\Sigma$ leads to the corresponding
multiplicities $M_2(y_1,k_1;y_2,k_2)$.
In the case when both $k_1\ll Q_s$ and $k_2\ll Q_s$
\beq
M_2=A^{4/9}\frac{3}{10}\gamma^2\frac{4}{\ln
2}\frac{\ba^2}{k_1^2k_2^2}\langle r^2\rangle _Pf^2(0)
e^{2\Delta Y-\epsilon(y_1+y_2)}\sqrt{\frac{\pi}{\beta y_1y_2 (2Y-y_1-y_2)}}.
\eeq
If both $k_1\gg Q_s$ and $k_2\gg Q_s$ then
\beq
M_2=A^{8/9}\frac{3}{14}\frac{4}{\ln
2}\frac{\ba^2}{k_1^3k_2^3}\langle r^2\rangle _P\lambda^2
\gamma^4e^{2\Delta
Y-\epsilon_1(y_1+y_2)}\sqrt{\frac{\pi}{\beta y_1^2y_2^2(2Y-y_1-y_2)}}.
\eeq
Finally in the case $k_1\gg Q_s$ and $k_2\ll Q_s$ we find
\beq
M_2=A^{2/3}\frac{1}{4}\frac{4}{\ln
2}\frac{\ba^2}{k_1^3k_2^2}\langle r^2\rangle _P\lambda f(0)\gamma^3
e^{2\Delta Y-\epsilon_1y_1-\epsilon y_2}\sqrt{\frac{\pi}{\beta y_1^2
y_2(2Y-y_1-y_2)}}.
\eeq

It is convenient to introduce a ratio
\beq
R(y_1,k_1;y_2,k_2)=\frac{M_2(y_1,k_1;y_2,k_2)}{M_1(y_1,k_1)M_1(y_2,k_2)},
\eeq
for which for all three limiting cases considered above we obtain  a simple
expression:
 \beq
R(y_1,k_1;y_2,k_2)=C\frac{1}{16\ln 2}\frac{\langle r^2\rangle _P}{R_P^2}
\sqrt{\frac{\beta (Y-y_1)(Y-y_2)}{\pi  (2Y-y_1-y_2)}},
\label{ccceq}
\eeq
where for the cases $k_1,k_2\ll Q_s$, $k_1,k_2\gg Q_s$ and $k_1\gg Q_s,k_2\ll Q_s$ the
coefficient $C$ is 32/15, 50/21 and 20/9 respectively. If the two produced
jets of hadrons are taken symmetric
in the c.m. system for $hN$ collisions with the rapidity distance $y $ then
\beq
y_1=\frac{1}{2}(Y+y), \ \ y_2=\frac{1}{2}(Y-y),
\eeq
and
\beq
R(y_1,k_1;y_2,k_2)=C\frac{1}{32\ln 2}\frac{\langle r^2\rangle _P}{R_P^2}
\sqrt{\frac{\beta}{\pi}}\,\Gamma(y_1,y_2),
\label{rsim}
\eeq
with
\beq
\Gamma(y_1,y_2)=\sqrt{Y-\frac{y^2}{Y}},\ \
\Gamma(y_1,y_1)=\sqrt{Y-y},\ \ 
\Gamma(y_2,y_2)=\sqrt{Y+y}.
\label{eqcoef}
\eeq
Eqs. (\ref{ccceq}), (\ref{rsim}) and (\ref{eqcoef}) are the central result of
this Section.

In terms of the ratio $R$ the correlation coefficient is given by
\beq
B=\frac{M_1(y_B,k_B)}{M_1(y_F,k_F)}\frac{R(y_F,k_F;y_B,k_B)-1}
{R(y_F,k_F;y_F,k_F)-1}.
\label{bratio}
\eeq
The coefficient in Eq. (\ref{rsim}) is very small e.g. $\sim 0.1$ for
$\alpha_s\simeq 0.2$. Thus both the numerator and denominator in this equation
are negative except for very large energies ($Y>100$ in the mentioned
example). Furthermore, $R$ may be smaller than one for some small window in
$y$ and $Y$, but it is generically larger than one. To illustrate it,
at large $Y$ and fixed $y$ we conclude from (\ref{rsim}) that $R$ is
independent of $y$ in all cases. So if $k_F$ and $k_B$ have the same order of
magnitude (either much smaller or much larger than $Q_s$) the second ratio in
(\ref{bratio}) is equal to unity.  In this case we have a simple result
\beq
B=\frac{M_1(y_B,k_B)}{M_1(y_F,k_F)}\,.  \eeq
For the most important case from
the practical point of view, $k_F,k_B\ll Q_s$,
we then find
\beq
B=\frac{k_F^{2}}{k_B^{2}}e^{\epsilon y},
\eeq
or for windows symmetric also in the phase volume of transverse momenta
simply
\beq
B=e^{\epsilon y}\,.
\label{bcoef}
\eeq
The concrete value of $\epsilon$
depends on the chosen value for $\alpha_s$. With a typical value $\alpha_s=0.2$
we find $\epsilon\simeq 0.1\div 0.15$.

It is not difficult to obtain predictions for $B$ for all other theoretically
possible cases. If both $k_F$ and $k_B$ are much larger than the saturation momentum 
we obtain the correlation coefficient (\ref{bcoef}) with $\epsilon$ substituted by
$\epsilon_1$. In still more exotic cases, when one of the momenta is much smaller
and the other much larger than $Q_s$, the second ratio in (\ref{bratio})
begins to depend on $Y$ non-trivially because of different coefficients $C$
in (\ref{rsim}). Also a nontrivial dependence on $A$ appears, due to different
powers of $A$ in (\ref{mlow}) and (\ref{mhigh}). The explicit formulas can be easily
written using  (\ref{mlow}), (\ref{mhigh}) and (\ref{rsim}).  We do not present them due
to a small probability of the corresponding experimental setup.

As we see in all cases the correlation coefficient turns out to be  different from the
probabilistic predictions. For symmetric windows  it is 
independent of $A$, generically
greater than unity and grows (rather slowly) with the
rapidity distance.

\section{Color Glass Condensate}

In this Section we follow the lines in \cite{levin,AMcLP}
to obtain an expression
for the correlation strength $B$ in hadron-nucleus, $pA$
collisions, considering the
hadron as a saturated object with some saturation scale $Q_{s,p}^2(y)<
Q_{s,A}^2(y)
\propto A^\delta$, $\delta>0$, as done in \cite{Dumitru:2001ux,Kharzeev:2002ei}.
For the multiplicity of produced gluons,
one gets in a small overlap area $a^2$ (with $a\ll R_0$
corresponding
to the correlation length of the classical fields) between
projectile and target:
\beq
\left\langle {dN \over dy}\right\rangle \sim {Q_{s,min}^2(y)\over \alpha_s\left(
Q_{s,min}(y)\right)}\, a^2,\ \ Q_{s,min}^2(y)={\rm min}\left\{
Q_{s,p}^2(y),Q_{s,A}^2(y)\right\}.
\label{mulcgc0}
\eeq
After integration over impact parameter one gets an overlap area $S$
i.e.
\beq
\left\langle {dN \over dy}\right\rangle \sim {Q_{s,min}^2(y)\over \alpha_s\left(
Q_{s,min}(y)\right)}\, S
\label{mulcgc}
\eeq
such that $S Q_{s,h}^2 \propto N_{part,h}$, the number of nucleons from hadron
$h$ participating in the collision ($N_{part,p} \equiv 1$),
see also \cite{Kharzeev:2002ei}.

The numerator in Eq. (\ref{nuclb}) (coming from diagrams like that in
Fig. 6 $a$)
results in
\beq
\left\langle {dN \over dy_F}{dN \over dy_B}
\right\rangle - \left\langle {dN \over
dy_F}\right\rangle \left\langle {dN \over dy_B}\right\rangle\sim
{Q_{s,min}^2(y_F)\over \alpha_s\left(
Q_{s,min}(y_F)\right)} {Q_{s,min}^2(y_B)\over \alpha_s\left(
Q_{s,min}(y_B)\right)}\, a^4.
\label{cgcunc0}
\eeq
%%%%%%%%%%%%%%%%%%%%%%%%%%%%%%%%%%%%%%%%%%%%%%%%%%%%%%%%%%%%%%%%%%%%
\begin{figure}[ht]
\centering
\includegraphics[width=12cm]{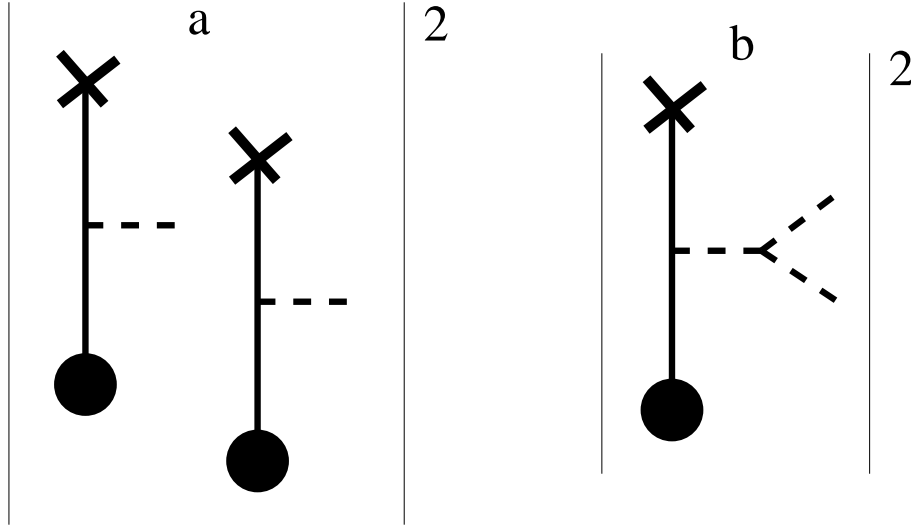}
\caption{Diagrams contributing to the double inclusive cross-section in the
CGC. The black dot and the cross correspond to the projectile and the
target from which the classical fields, represented by the solid lines,
come. The dashed lines represent the emitted gluons.}
\label{Fig6}
\end{figure}
%%%%%%%%%%%%%%%%%%%%%%%%%%%%%%%%%%%%%%%%%%%%%%%%%%%%%%%%%%%%%%%%%%%%
Now, for the integration over impact parameter we consider that
$a^2\sim 1/Q_{s,max}^2(y_F,y_B)$, $Q_{s,max}^2(y_F,y_B)={\rm max}\left\{
Q_{s,p}^2(y_F),Q_{s,A}^2(y_F),Q_{s,p}^2(y_B),Q_{s,A}^2(y_B)\right\}$,
which results in
\beq
\left\langle {dN \over dy_F}{dN \over dy_B}
\right\rangle - \left\langle {dN \over
dy_F}\right\rangle \left\langle {dN \over dy_B}\right\rangle\sim
{Q_{s,min}^2(y_F)\over \alpha_s\left(
Q_{s,min}(y_F)\right)} {Q_{s,min}^2(y_B)\over \alpha_s\left(
Q_{s,min}(y_B)\right)}{S\over Q_{s,max}^2(y_F,y_B)}\,. 
\label{cgcunc}
\eeq
For the denominator in Eq. (\ref{nuclb}) there is an additional piece coming
from diagrams like that in Fig. 6 $b$,
\beq
\left[\left\langle {dN \over dy_F}{dN \over dy_B}
\right\rangle - \left\langle {dN \over
dy_F}\right\rangle \left\langle {dN \over dy_B}\right\rangle\right]^\prime\sim
Q_{s,min}(y_F) Q_{s,min}(y_B) e^{-\kappa (y_F-y_B)} S.
\label{cgcuncp}
\eeq
As discussed in \cite{AMcLP}, this piece is ${\cal O}(\alpha_s^2)$ suppressed
compared to (\ref{cgcunc}).
It contains an exponential damping factor, with $\kappa \sim 1$, which
motivates its absence in the numerator for large enough $y_F-y_B$.
Also note that for symmetric $AA$ collisions,
both (\ref{cgcunc}) and (\ref{cgcuncp}) are ${\cal
O}(A^{-2/3})$ suppressed
compared to the square of (\ref{mulcgc}), as discussed in \cite{levin}.

Neglecting any possible interference between these two kinds of diagrams and
for large enough $y_F-y_B$,
the final expression for Eq. (\ref{nuclb}) reads\footnote{An error in this
formula in a previous version of this manuscript, led
to wrong conclusions in this
Section. We warmly thank J.G.Milhano for pointing it out.}
\beq
B= {\alpha_s\left(Q_{s,min}(y_F)\right) Q_{s,max}^2(y_F,y_F)
Q_{s,min}^2(y_B) \over \alpha_s\left(Q_{s,min}(y_B)\right)
Q_{s,max}^2(y_F,y_B)
Q_{s,min}^2(y_F)}\,
\left[ 1+c\,\alpha_s^2\left(Q_{s,min}(y_F)\right)
{Q_{s,max}^2(y_F,y_F)
\over Q_{s,min}^2(y_F)}
\right]^{-1},
\label{cgcnuclb}
\eeq
where $c$ is a constant. Note that for $AA$ collisions and symmetric intervals
$y_B=-y_F$, this expression reduces to that found in \cite{AMcLP},
$B=\left[1+c \alpha_s^2\right]^{-1}$,
from which the
correlation strength was argued to increase with centrality and energy of the
collision. For the asymmetric case Eq. (\ref{cgcnuclb}), these behaviors
depend on
the considered rapidities $y_F,y_B$.

To illustrate this,
let us take the parametrizations \cite{Kharzeev:2002ei}
\beq
Q_{s,p}^2(y)=Q_0^2 \left( {E_{cm}\over Q_0}\right)^\lambda e^{-\lambda y},
\ \ Q_{s,A}^2(y)=Q_{s,p}^2\left(-y, Q_0^2\to
%Q_0^2 \left[ {N_{part}\over 2}\right]^{1/3}\right),
Q_0^2 A^{1/3}\right),
\label{qs}
\eeq
with $Q_0^2$ a constant with dimension of momentum squared,
$E_{cm}$ the collision energy in the center-of-mass system,
%$N_{part}$ the number of participants in the
%collision,
$\lambda \sim 0.3$,
and rapidities defined in the center-of-mass system in which
the projectile is located at $y=Y/2$ and the nuclear target at $y=-Y/2$.
The solution of the equation
\beq
Q_{s,p}^2(y_c)=Q_{s,A}^2(y_c)
\label{ycrit}
\eeq
defines \cite{Kharzeev:2002ei}
a critical rapidity $y_c\simeq-4\div -3<0$ such that for $y<y_c$,
$Q_{s,min}^2(y)=Q_{s,A}^2(y)$, while for $y>y_c$,
$Q_{s,min}^2(y)=Q_{s,p}^2(y)$. Let us examine several situations:
\begin{itemize}
\item
For $y_c<y_B=-y_F$ - the most feasible situation from the experimental point
of view, 
$Q_{s,min}(y_F)=Q_{s,p}(y_F)$, $Q_{s,min}(y_B)=Q_{s,p}(y_B)$ and
$Q_{s,max}(y_F,y_B)=Q_{s,A}(y_F)$, the correlation strength $B$ is generically
greater
than 1 and decreases with
increasing $A$. It shows a non-monotonic behavior with increasing energy,
first increasing and then decreasing towards a limiting value
$e^{2 \lambda y_F}>1$ independent of $A$.
This case coincides with the one in (\ref{bcoef}) (with $\epsilon
\leftrightarrow \lambda$ and $y \leftrightarrow 2y_F$), which can be easily
understood as the consideration of a dilute projectile in the previous Section
is equivalent to the
limit $y_c \to -\infty$ here.
\item
For $y_B=-y_F<y_c$, 
$Q_{s,min}(y_F)=Q_{s,p}(y_F)$, $Q_{s,min}(y_B)=Q_{s,A}(y_B)$ and
$Q_{s,max}(y_F,y_B)$ $=Q_{s,A}(y_F)$,  the correlation strength $B$ is
generically greater
than 1 and increases with
increasing $A$. Again, it shows a non-monotonic behavior with increasing energy,
first increasing and then decreasing towards a limiting value
$A^{(2-\lambda)/6}>1$ independent of $y_F=-y_B$.
\item For
$y_B<y_F<y_c$ (a situation within the kinematical reach of the Large Hadron
Collider (LHC)
with e.g.
$y_B=-7$ and $y_F=-5$), $Q_{s,min}(y_F)=Q_{s,A}(y_F)$,
$Q_{s,min}(y_B)=Q_{s,A}(y_B)$ and
$Q_{s,max}(y_F,y_B)=Q_{s,p}(y_B)$, the correlation strength $B$ is generically
smaller
than 1 and
increases with
increasing $A$. Opposite to the two previous cases, it shows a monotonic
increase with increasing energy towards a limiting value $e^{-2 \lambda
(y_F-y_B)}$ $<1$ independent of $A$.
\end{itemize}
Let us note that these estimates have been done for truly asymptotic values of
energy and nuclear size and for a concrete choice of the parametrization for
the
saturation scale, the behavior for the experimentally accessible
situation depending on this concrete choice
and on the value of parameter $c$. Nevertheless, the fact
that the correlation strength is larger or smaller than one is generic.

The results in this Section
should be applicable for transverse momenta of the order or
smaller than the corresponding saturation scales.
In any case, the correlation strength is larger than 1 for the most feasible
experimental situation of forward and backward rapidity windows symmetric
around the center-of-mass, a behavior
which coincides to that generically
found in the previous Section, see e.g. (\ref{bcoef}).

\section{Conclusions}

In this paper we analyze the
long-range rapidity correlations in hadron-nucleus collisions.
First we recall
the standard results in the probabilistic model. The correlation strength $B$
shows an increasing behavior with increasing nuclear size or centrality,
tending to unity for $A\to \infty$.

Next we turn to
perturbative QCD. We consider interacting BFKL
pomerons in the form of fan diagrams coupled to a dilute projectile. After
examining the required single and double inclusive cross-sections
\cite{braun3,braun4}, analytic
estimates are done for very large rapidities due to the difficulties for a
complete consideration of the double inclusive density.
The correlation strength results
weakly depending on energy and centrality or nuclear size, and generically
greater than unity.
Note that these results are rigorously applicable to DIS, but require certain
caution if applied to $hA$ scattering with ordinary hadrons.

Finally, we turn to the Color Glass Condensate framework.
Taking the projectile as a saturated object characterized by a saturation
scale smaller - as expected -
than that of the nucleus, we extend to asymmetric collisions
the analysis done in \cite{levin,AMcLP} for
nucleus-nucleus collisions. For the most feasible
experimental situation of forward and backward rapidity windows symmetric
around the center-of-mass, the resulting correlation
strength turns out to be larger than unity and
shows a non-monotonic behavior with increasing energy, first increasing and
then decreasing to a limiting value. Its behavior with
increasing centrality or nuclear size depends on the considered
rapidity
windows.

A note of caution is in order.
The correlations considered in this paper are those coming from particle
production in the initial stage of the collision. Subsequent stages may modify
the predicted
behaviors.
In any case, hadron-nucleus collisions should offer a more
reliable setup than nucleus-nucleus in this respect, as the production of a
dense, eventually
thermalized medium is not expected. Besides, hadronic rescattering of
the produced secondaries is expected to play a little role except in the
region close to the rapidity of the nucleus.
With all these caveats in mind,
phenomenological applications of our results to $dAu$ collisions at the
Relativistic Heavy Ion Collider and
$pA$ collisions at the LHC are the obvious extension of this work.

\section{Acknowledgments}
MAB has been financially
supported by grants RNP 2.1.1.1112 and RFFI 06-02-16115a
of Russia, and NA and CP by Ministerio de Educaci\'on y Ciencia of
Spain under project FPA2005-01963 and by Xunta
de
Galicia (Conseller\'{\i}a de Educaci\'on). NA also acknowledges financial
support by Ministerio de Educaci\'on y Ciencia of
Spain under a contract Ram\'on y Cajal. We thank J.Dias de Deus,
F.Gelis, L.McLerran,
A.H.Mueller and
B.Srivastava for useful discussions.
Special thanks are due to J.G.Milhano who pointed us an error in
(\ref{cgcnuclb}) in an earlier version of this paper.
MAB also thanks Departamento
de F\'{\i}sica de Part\'{\i}culas of the Universidade de Santiago de
Compostela for warm hospitality.

\end{document}